\documentclass[aps,groupedaddress,showpacs,%
amsmath,preprintnumbers,10pt]{revtex4}

\usepackage{graphicx}
\usepackage{amsfonts}
\usepackage{amssymb}
\usepackage{amsbsy}
\usepackage{amsmath}
\usepackage{latexsym}
\usepackage{bm}
\newcommand*{\di}{\partial}

\newcommand*{\KG}{\text{\tiny{KG}}}
\newcommand*{\tot}{\text{\tiny{tot}}}
\newcommand*{\poly}{\text{poly}}
\renewcommand*{\pl}{\text{pl}}
\newcommand*{\M}{M_\star}
\newcommand*{\D}{\mathrm{D}}

\def\p {\partial}

\def\be {\begin{equation}}
\def\ee  {\end{equation}}
\def\bea {\begin{eqnarray}}
\def\eea {\end{eqnarray}}
\def\nn {\nonumber}

\begin{document}
\preprint{ }
\title{Background independent quantization and wave propagation}
\author{Golam Mortuza Hossain}
\email{ghossain@unb.ca}
\author{Viqar Husain}
\email{vhusain@unb.ca}
\author{Sanjeev S.~Seahra}
\email{sseahra@unb.ca } \affiliation{ Department of Mathematics and
Statistics, University of New Brunswick, Fredericton, NB, Canada E3B
5A3} 
\date{June 22, 2009}
\begin{abstract}

We apply a type of background independent ``polymer'' quantization
to a free scalar field in a flat spacetime.  Using semi-classical
states, we find an effective wave equation that is both nonlinear
and Lorentz invariance violating. We solve this equation
perturbatively for several cases of physical interest, and show that
polymer corrections to solutions of the Klein-Gordon equation depend
on the \emph{amplitude} of the field. This leads to an effective
dispersion relation that depends on the amplitude, frequency and
shape of the wave-packet, and is hence distinct from other modified
dispersion relations found in the literature. We also demonstrate
that polymer effects tend to accumulate with time for
plane-symmetric waveforms. We conclude by discussing the possibility
of measuring deviations from the Klein-Gordon equation in particle
accelerators or astrophysical observations.

\end{abstract}

\maketitle

\section{Introduction}\label{sec:intro}
There has been much interest in the past few years on possible
quantum gravitational effects on wave propagation. The basic idea is
that if the metric is quantized, effective wave dynamics follow from
replacing the classical metric in the wave equation by the
expectation value of a metric operator in a suitable quantum state.
Such a state would provide a ``quantum gravity corrected''
background spacetime with a built in fundamental discreteness scale.
It has been conjectured that this may lead to Lorentz symmetry
violating modifications to conventional wave propagation and
dispersion relations \cite{lqg-wave}.  Phenomenological
effects have been studied by introducing explicit Lorentz violating terms 
in particle physics models \cite{myers-posp} with potentially significant effects 
\cite{collins}  that could require fine tuning to agree with some observations. 

Several authors have previously considered the effects of dispersion
relations incorporating high energy corrections on Hawking radiation
\cite{unruh-disp, ted-disp} and inflationary cosmology
\cite{brand-disp}.  In the studies, the precise form of the
corrections was specified on an \emph{ad hoc} basis.  It was found
that the Hawking effect is not affected by such modifications,
whereas in cosmology some classes of dispersion relations do affect
the spectrum of quantum fluctuations.

It has been argued that quantum gravity modifications to the group
velocity of wave-packets can lead to potentially observable effects
on the signal from gamma ray bursts \cite{gamma-exp}.  Observed
violations of the GZK cutoff \cite{GZK} in the spectrum of high
energy cosmic rays can also indicate a break-down of Lorentz
symmetry at high energy.  Attempts have been made to systematically
derive constraints on parameters appearing in modified dispersion
relations such as
\be \label{DSRDispersion}
 E^2 =m^2 + p^2 \left[1 + \zeta \left(
\frac{p}{p_{\pl}}\right)^\gamma \right] + \cdots
 \ee
where $E$, $m$ and $p$ denote the energy, mass and momentum
respectively; $p_{\pl}$ is the Planck momentum and $\zeta$ and
$\gamma>0$ are real parameters \cite{ted-review}.

It is important to ask which ideas associated with the the
construction of a quantum theory of gravity lead to potentially
observable effects on wave propagation in the semi-classical regime.
The work in this direction utilizes a background independent
quantization scheme for the dynamical variables associated with
geometry. This approach to quantum theory may be applied to any
system, and in particular also to matter degrees of freedom. Indeed
for  a complete quantization of a gravity-matter system it is
natural to apply the same procedure to all degrees of freedom.

Our purpose in this paper is to study the effects of a background
independent quantization on wave propagation {\it without any
quantization of geometry}.  The  approach may be viewed as the study
of polymer quantum fields on curved spacetime, which is a general
area of interest in its own right.

The simplest case is the massless  scalar field on a Minkowski
background. The kinematics of this quantization, and its comparison
to the usual Fock procedure, has been discussed at a mathematical
level \cite{Thiemann:1997rt,qsd-n,als-fock} using a direct
application of the ideas developed in loop quantum gravity (LQG).
However, so far there appears to have been relatively little study
of possible physical consequences of this approach. This is what we
undertake here, but with a number of essential differences in
approach on which we will elaborate.

This paper is organized as follows. In \S\ref{sec:quantization}, we
review some aspects of background independent quantization, with
emphasis on a new set of basic variables for quantization. (These
have application well beyond the scenario addressed here.)  We then
introduce a class of semi-classical states, and show how these lead
to a modified wave equation that is both nonlinear and Lorentz
invariance violating.  This has obvious and significant consequences
for wave propagation, even without considering any quantum gravity
effects. It provides a sharp contrast to the linear modifications of
the wave equation used in all the previous applications mentioned
above.  We then describe a perturbative solution scheme for the
semi-classical wave equation in \S\ref{sec:perturbative}, and its
specialization to situations with planar and spherical symmetry. In
\S\ref{sec:phenomenology} we discuss the phenomenological
implications of the model in the context of modified dispersion
relations, the coherence length of collimated beams (with
application to the Large Hadron Collider and high energy cosmic
rays), and pseudo-spherical radiation from astrophysical sources
(such as gamma-ray bursts).  A summary and discussion of our results
is in \S\ref{sec:discussion}.

\section{Quantization and the effective wave
equation}\label{sec:quantization}

\subsection{Polymer formalism}
We consider a massless free scalar field in four spacetime
dimensions with the following background metric
\be g_{\mu\nu} dx^{\mu}dx^{\nu} = -dt^2 + q_{ab}dx^a dx^b ~. \ee
Here $g_{\mu\nu}$ is the spacetime metric whereas $q_{ab}$ denotes
the spatial metric. The canonical phase space  variables are
$(\phi,P_\phi)$ with scalar matter Hamiltonian
\be H = \int d^3x \left( \frac{1}{2\sqrt{q}} P_\phi^2 +
\frac{\sqrt{q}}{2} q^{ab}\p_a \phi\p_b\phi \right), \ee
where $ q = det(q_{ab})$. The  usual wave equation for scalar field
follows from this via Hamilton's equations.

Let us consider the functions
\be \phi_f = \int d^3y \sqrt{q} f(y) \phi(y), \ \ \ \ \ \ \ \
U_\lambda(P_\phi)= \exp\left[i\lambda P_\phi/\sqrt{q}\ \right],
\label{basic-obs} \ee
as the new set of basic variables that are to be realized as
operators in the quantum theory. Here  $f(x)$ is a scalar  and
$\lambda$ is a  real constant with the dimension of \emph{length
squared} in natural units. The factors of $\sqrt{q}$ are necessary
to balance density weights in the integral and in the exponent.
(These variables may be viewed as  the ``dual''  \cite{hw-sing} to
those used  in the polymer quantization of a particle system
\cite{polymer, halvorson} motivated by loop quantum gravity LQG
\cite{Thiemann:2001yy,Ashtekar:2004eh,Smolin:2004sx,Rovelli:2008zza},
where  the exponentiated configuration variable is used as the new
variable. The dual for quantum mechanical systems has been discussed
in  \cite{halvorson}.) We note however that there are important
differences with the variables used for scalar field quantization in
\cite{als-fock}, which do not use the available background metric in
their definition.
Our variables satisfy the canonical Poisson bracket
\be \{ \phi_f , U_\lambda(P_\phi(x)) \}  = i \lambda f(x)
U_\lambda(P_\phi(x)) ~. \label{basicpb} \ee
A localized field  may be defined by taking for example $f(x)$ to be
a Gaussian \be G(x,x_k,\beta)= e^{-{\beta^2}(x-x_k)^2} ~, \ee which
is sharply peaked ($1 \ll {\beta^2}$)  at a point $x_k$. We   assume
this in the following and write $\phi_G(x_k)\equiv \phi_k.$

Since the representation of these variables that we will use is
based on that  for polymer quantum mechanics \cite{polymer,
halvorson}, we briefly review this quantization before generalizing
it to field theory.  The Hilbert space  is the space of almost
periodic functions, where a wave function is written as the linear
combination
\be \psi(p) = \sum_{k=1}^{N} c_k e^{ix_k p} = \sum_{k=1}^{N} c_k
\langle p|x_k\rangle~. \ee
Here, the set of points $\{x_k\}$ is a selection (graph) from the
real line.  The inner product is
\be \langle x| x'\rangle = \lim_{T\to\infty} \frac{1}{2T}\int_{-T}^T
dp\  e^{-ipx}e^{ipx'} = \delta_{x,x'} , \ee
in which plane waves are normalizable (the right hand side is the
generalization of the Kronecker delta to continuous indices). On
this Hilbert space  the configuration and translation operators act
as
\be -i\frac{\p}{\p p} e^{i p x_k} = x_k e^{i px_k}, \ \ \ \ \ \ \
\widehat{e^{i\lambda p}} e^{ipx_k} = e^{ip(x_k +\lambda)}. \ee
The momentum operator is not defined on this Hilbert space.
Consequently, only the \emph{finite} translation can be realized
rather than the \emph{infinitesimal} translation.

The generalization of this quantization to field theory is
straightforward.  The Poisson bracket of the basic scalar field
variables is  realized as an operator relation on a Hilbert space
with basis states
\be | a_1,a_2,\cdots ,a_n\rangle \ee
where the real numbers $a_i$ represent scalar field values at the
spatial lattice points $x_i$. The inner product  is
\be \langle a_1',a_2',\cdots ,a_n'| a_1,a_2,\cdots ,a_n\rangle=
\delta_{a_1',a_1}\cdots \delta_{a_n',a_n} \ee
when two states are associated with the same set of  lattice points;
if not the inner product is zero.  For our purpose it suffices to
consider a fixed spatial lattice of the type used in numerical
computation.
This inner product is background (metric) independent in the same
way  as for example that for  Ising model spins; the difference is that
for the latter there is a finite dimensional spin space at each
lattice point, rather than a real number representing a field value
at each point. In contrast the usual Fock space quantization uses
the metric dependent Klein-Gordon inner product. 

The configuration and translation operators are defined by the
following action:
\be \hat{\phi}_f| a_1,a_2,\cdots a_n\rangle\equiv \sum_i a_i f(x_i)|
a_1,a_2,\cdots a_n\rangle, \ee
\be \widehat{U}_\lambda(P_\phi(x_k))| a_1,a_2,\cdots a_n\rangle
\equiv | a_1,a_2,\cdots , a_k+ \lambda, \cdots a_n\rangle. \ee
It is readily verified that the commutator of these operators is a
faithful realization of the corresponding Poisson bracket.  The
parameter $\lambda$ represents the discreteness scale in field
configuration space.
In this representation the momentum operator does not exist. There
is however an alternative $\lambda$ dependent definition of
`momentum' given by
\be \hat{P}_\phi^\lambda(x_k) \equiv \frac{\sqrt{q}}{2i\lambda}\
\left[\hat{U}_\lambda(x_k) - \hat{U}_\lambda^\dagger (x_k) \right],
\label{mom} \ee
which can be used to define the Hamiltonian operator.

\subsection{Semi-classical states}\label{sec:semi-classical}

Our goal is to obtain an effective Hamiltonian using this
quantization procedure. This will be done by using semi-classical
Gaussian states peaked at  a classical phase space configuration.
Such states have  been defined for Friedmann-Robertson-Walker
cosmology  in the same representation \cite{hw-cosm}, and we use
those results here with a suitable generalization to field theory.

For definiteness, we consider background spacetime here to be
Minkowski spacetime with $\sqrt{q} = 1$. This is also the focus of
the present paper.
In this spacetime  consider a space lattice point $x_k$ and the
basis states $|a_k=m\lambda\rangle_k$ at this point, where $m$ is an
integer. A semi-classical state at $x_k$ is defined by
\be
 |P_0,\phi_0\rangle^{\sigma,\lambda}_{x_k} =
\frac{1}{C}\sum_{m=-\infty}^\infty e^{-(\lambda m)^2/2\sigma^2}
e^{m\lambda \phi_0} e^{-im\lambda P_0 }|m\lambda \rangle_{x_k}.
 \label{semiclass}
\ee This is a Gaussian  state of width $\sigma$ where the (real)
parameters $P_0$ and $\phi_0$ represent field values corresponding
to a classical configuration at the point $x_k$.
The normalization constant $C>0$ is given by the convergent sum
\be C^2=\sum_{m=-\infty}^\infty e^{-\lambda^2m^2/\sigma^2}
e^{2\phi_0 \lambda m}. \label{norm} \ee
Calculation with this  state gives the following expectation value
\cite{hw-cosm}
\be \langle \hat{U}_\lambda \rangle =  e^{i\lambda P_0(x_k)}\
e^{-\lambda^2/4\sigma^2} \ K(\lambda,\sigma,\phi_0) ~,
\label{expecU} \ee
where
\be K(\lambda,\sigma,\phi_0)=\left(\frac{  1+ 2\sum_{m\ne
0}\cos\left[ \frac{2\pi m\phi_0\sigma^2}{\lambda }\left(1+
\frac{\lambda}{2\phi_0\sigma^2}\right) \right]
e^{-\pi^2m^2\sigma^2/\lambda^2} }{ 1+ 2\sum_{m\ne 0}\cos\left[
\frac{2\pi m\phi_0\sigma^2}{\lambda} \right]
e^{-\pi^2m^2\sigma^2/\lambda^2} }\right) ~. \ee
Equation (\ref{expecU}) together with the definition (\ref{mom})
gives the expectation value
\be \langle \hat{P}_\phi^\lambda(x_k)\rangle = \frac{\sin[P_0(x_k)
\lambda]}{\lambda}\ e^{-\lambda^2/4\sigma^2}\
K(\lambda,\sigma,\phi_0). \ee
This formula has the limits
\bea
 \lim_{\sigma\rightarrow \infty}\ \langle \hat{P}_\phi^\lambda(x_k)\rangle
&=& \sin[ P_0(x_k)\lambda ]/\lambda, \\
 \lim_{\lambda\rightarrow 0}\ \langle \hat{P}_\phi^\lambda(x_k)\rangle
&=& P_0(x_k).
 \eea
The first limit {\em i.e.} ignoring the width corrections of the
quantum states, shows that the semi-classical state on the field
lattice is peaked at the corresponding phase space value. The second
limits shows that in the field continuum limit of the momentum
expectation value has the appropriate classical value. This is
crucial for the quantization scheme to be a viable quantization
method given only \emph{finite} field translation operators exist in
the representation we are using.

\subsection{Effective Hamiltonian}\label{sec:Hamiltonian}

We now turn to derive an effective Hamiltonian for the scalar field
using the semi-classical states (\ref{semiclass}).
Let $\mathcal{H}(P_\phi(x),\phi(x))$ be the Hamiltonian density and
we define the effective Hamiltonian density $\mathcal{H}^{\rm eff}$
by the expectation value
\be \mathcal{H}^{\rm eff} (P_0(x_k),\phi_0(x_k);\sigma,\lambda)
\equiv \frac{1}{2 }\ \langle P_0,\phi_0| ( \hat{P}_\phi^\lambda) ^2
%\over {\sqrt{q}}}
+  \widehat{\nabla^2\phi}   |P_0,\phi_0
\rangle_{x_k}^{\sigma,\lambda} ~. \ee
The kinetic part of the operator gives
\bea \langle P_0,\phi_0|  (\hat{P}_\phi^\lambda)^2  |P_0,\phi_0
\rangle_{x_k}^{\sigma,\lambda} &=& {1 \over  \lambda^2}\ \langle
P_0,\phi_0|  \left( 2- \hat{U}_{2\lambda} (P_\phi(x_k))
 -  \hat{U}_{2\lambda}^\dagger (P_\phi(x_k)) \right)
|P_0,\phi_0 \rangle_{x_k}^{\sigma,\lambda}\nn\\
&=&  {1 \over \lambda^2}\ \sin^2\left(  P_0 \lambda \right) + {\cal
O}\left(\frac{1}{\sigma}\right), \eea
where the last term represents the width corrections of the
semi-classical states.
The scalar gradient operator can be defined by an expression such as
 \be
   \frac{\p\phi}{\p x} \rightarrow   \left(\frac{1}{\epsilon} \hat{\phi}_{x_k+\epsilon} - \hat{\phi}_{x_k}\right),
 \ee
 where $\hat{\phi}_{x_k} $ is defined with a Gaussian peaked at $x_k$  as described above, and
 $\epsilon$ is the spacing of a uniform lattice. (Note that there are two lattices in this formulation, one in field space  with spacing $\lambda$ and the other in physical $3-$space with spacing $\epsilon$.)
Thus  for wavelengths large compared to $\epsilon$  the  space
continuum approximation applies, and we can write
\be \mathcal{H}^{\rm eff} (x) = \frac{1}{2\lambda^2}
\sin^2\left(P\lambda\right) + \frac{1}{2}\ \nabla^2\phi, \ee
where we have dropped the subscripts on $P_0$ and $\phi_0$. In this
expression and what follows, we have neglected corrections due to
the finite width of the semi-classical state.  That is, we will work
in the $\sigma \rightarrow \infty$ limit for the rest of this paper.
Finally, the effective Hamiltonian
\be H^{\rm eff} = \int d^3x \  \mathcal{H}^{\rm eff}(P,\phi) \ee
gives the modified wave equation
\begin{equation}
    \frac{\di^2\phi}{\di t^2} - \left[ 1-4\lambda_{\star}^2\left( \frac{\di\phi}{\di t} \right)^2
    \right]^{1/2} \nabla^2 \phi = 0,
    \label{waveeqn}
\end{equation}
where we have set $\lambda =\lambda_{\star}$ as a fixed
``polymerization scale''. The nonlinear term represents the
continuum remnant  of polymer quantization. This equation is not
Lorentz invariant (due to the $\dot{\phi}^2$ term), but is
  invariant under Galilean and parity ($x\rightarrow -x,\ t\rightarrow -t$)
transformations, as well as field reflection, $\phi\rightarrow
-\phi$.  We note also that  the effective Hamiltonian is the integral of local
density so local casual evolution via Hamilton's equations is assured. Indeed
the corrections to the Hamiltonian in an expansion in $\lambda_*$ gives terms
that are higher derivative in momenta, which are like higher derivative terms from
the lagrangian perspective.

A generalization of this equation to arbitrary space metric
$q_{ab}$ with lapse $N=1$ and shift $N^a=0$ derived using the same
considerations appears in Appendix \ref{app:wave equation on a
curved background}.

\subsection{Homogeneous scalar field on a cosmological
background}\label{sec:cosmology}
The formalism of polymer quantization presented here can be applied
to different background spacetimes. In particular, it is useful to
see what happens if we consider a reduction of the theory to the
homogeneous spacetime where it is a quantum mechanical system.
Consider the Friedmann-Robertson-Walker (FRW) background \be ds^2 =
-dt^2 +a(t)^2 (dx^2 + dy^2 +dz^2), \ee so that $\sqrt{q} = a^3$
where $a$ is the scale factor. The basic observables
(\ref{basic-obs}) therefore reduce to \be \phi_f = V_0 a^3 \phi,\ \
\ \ \ \ \ \ U_\lambda = e^{i\lambda P/a^3}, \ee where $V_0$ is a
fiducial volume obtained from the spatial integration, and we have
set $f=1$ (which is natural given that any other non-zero constant
can be absorbed in $V_0$.   The algebra obtained from the
fundamental bracket  (\ref{basicpb}) gives \be \{ \phi , U_\lambda
\} = {i \lambda\over V_0 a^3} U_\lambda. \ee

These relations have an interesting feature when  compared to the so
called ``improved dynamics'' \cite{improved-dyn} scenario in loop
quantum cosmology (LQC) \cite{Bojowald:2008zzb}: The appearance of
the $a^3$ factor in $U_\lambda$ and the reduced Poison bracket is
natural from the start, given the definition  of the basic variables
that we are using. This feature leads to the so-called dynamical
lattice refinement in cosmology that is necessary to reproduce
correct classical limit.  In the LQC context on the other hand,
there is no natural way to see how these factors can arise from the
full theory upon reduction of a holonomy, which has no metric
($\sqrt{q}$) dependent factors in it.

Using these variables one can study the effective dynamics of a
homogeneous polymer scalar field coupled to an FRW background, with
rather surprising results: there is singularity avoidance and a
built-in inflationary phase, {\it without any need to quantize
gravity} \cite{polyfrw}. In this scenario, the non-singular universe
is past inflating. This is in contrast to LQC where the singularity
avoidance is realized through a quantum \emph{bounce}  that occurs
when the density of matter reaches the Planck density
\cite{Bojowald:2001xe,Date:2004fj,improved-dyn}. Consideration of
inhomogeneous fluctuations in this inflating background can provide
a probe to study the polymer effects through the power spectrum of
cosmological fluctuations. This is presently under investigation.

\section{Perturbative solutions}\label{sec:perturbative}

\subsection{General solution}

Our goal in this section is to develop perturbative solutions to the
equation (\ref{waveeqn}). As a first step, we rewrite the equation
in terms of dimensionless coordinates and fields. Noting that the
parameter $\lambda_{\star}$ has dimensions of length squared, we
define a mass scale $\M$ by
\begin{equation}
    \M = 1/\sqrt{\lambda_{\star}} ~.
\end{equation}
Using the mass scale $\M$, we can define a set of dimensionless
quantities as follows
\begin{equation}\label{DimLessVars}
    \Phi = \phi/\M, \quad T = \M t, \quad X^i = \M x^i.
\end{equation}
In terms of these variables, the equation (\ref{waveeqn}) becomes
\begin{equation}
    \ddot\Phi - ( 1-4\dot\Phi^2
    )^{1/2} \D^2 \Phi = 0,
    \label{dimensionless wave}
\end{equation}
where an overdot indicates $\di/\di T$ and $\D_i = \di/\di X^i$.
Notice that neither $\lambda_{\star}$ nor $\M$ appears in this
equation.
The nonlinear wave equation (\ref{dimensionless wave}) admits the
trivial solution $\Phi = 0$.  We construct a perturbative series
about this exact solution as follows:
\begin{equation}\label{eq:perturbative series}
    \Phi = \epsilon \Phi_1 + \epsilon^2 \Phi_2 + \cdots = \sum_{i=1}^\infty \epsilon^i \Phi_i,
\end{equation}
where $\epsilon \ll 1$ is a small dimensionless parameter. If we put
this ansatz into the wave equation, expand in powers of $\epsilon$,
and set the coefficient of each power equal to zero, we obtain:
\begin{subequations}\label{eq:perturbative EOMs}
\begin{eqnarray}
    \label{eq:first order} \Box \Phi_1 & = & 0, \\
    \Box \Phi_2 & = & 0, \\
    \label{eq:third order} \Box \Phi_3 & = & 2\dot\Phi_1^2 \D^2\Phi_1, \\
    \Box \Phi_4 & = & 2\dot\Phi_1 ( \dot\Phi_1 \D^2 \Phi_2 + 2
    \dot\Phi_2 \D^2 \Phi_1), \\
    \Box \Phi_5 & = & 2(\dot\Phi_1^4 + \dot\Phi_2^2 + 2 \dot\Phi_1\dot\Phi_3)\D^2\Phi_1
    + 2\dot\Phi_1(\dot\Phi_1 \D^2 \Phi_3 + 2\dot\Phi_2\D^2
    \Phi_2),\\
    \nonumber & \vdots &
\end{eqnarray}
\end{subequations}
where $\Box = -\di_T^2 + \D^2$.

Before moving on, a few comments about these perturbative solutions
are in order. First, note that the perturbative equations of motion
(\ref{eq:perturbative EOMs}) imply that we may absorb the $\Phi_2$
contribution into $\Phi_1$ by making the substitution $\Phi_1 +
\epsilon \Phi_2 \rightarrow \Phi_1$.  That is, we can effectively
set $\Phi_2 = 0$.  This then implies that we can absorb the $\Phi_4$
into $\Phi_3$ via $\Phi_3 = \Phi_3 + \epsilon \Phi_4$, and so on.
Thus at the end, we are left with $\Phi_{2i} = 0$ and a perturbative
series with only odd powers of $\epsilon$.

Secondly, it is perhaps useful to elaborate on the nature of our
perturbative expansion.  In field theory, perturbative series
solutions often make use of dimensionless factors appearing in the
action or equations of motion as expansion parameters.  The
prototypical example of this is a scalar field with a
$V(\phi)=\mu\phi^4$ self-interaction.  In such a model, it is
natural to develop a perturbative solution in terms of the
dimensionless parameter $\mu$; i.e., $\phi = \sum_n \mu^n \phi_n$.
Our situation is somewhat different since the wave equation
(\ref{dimensionless wave}) has no small dimensionless parameters
appearing in it.  Hence, we instead develop a series solution in
terms of the \emph{amplitude} of $\Phi$.  Such expansions are
familiar from elementary physics; for example, consider the equation
of motion of a simple pendulum of length $\ell$ in a uniform
gravitational field $g$:
\begin{equation}
    \ddot\theta = - \sin\theta,
\end{equation}
where the dimensionless time is $T \equiv (g/\ell)^{1/2} t \equiv
\omega t$ and $\theta$ measures the angular position. Like
(\ref{dimensionless wave}), this equation has no naturally small
parameters appearing in it.  To solve for $\theta$, we employ an
expansion
\begin{equation}
    \theta(T) = \sum_{n=1}^\infty \epsilon^n \theta_n(T),
\end{equation}
which leads to equations of motion
\begin{equation}
    \ddot\theta_1 + \theta_1 = 0, \quad \ddot\theta_2 + \theta_2 = 0,
    \quad \ddot \theta_3 + \theta_3 = \tfrac{1}{6}\theta_1^3,
\end{equation}
and so on.  These are very similar in structure to
(\ref{eq:perturbative EOMs}), and we see that $\theta_1$ satisfies
the familiar equation of motion for a simple harmonic oscillator.
The parameter $\epsilon$ has a natural interpretation when we write
down the full $\theta$ solution
\begin{equation}
    \theta = \epsilon \cos(\omega t + \delta) + \cdots
\end{equation}
That is, $\epsilon$ is just the dimensionless amplitude of the
leading order term, which implies that the perturbative series is
only valid for small angle deviations from the vertical. In our
perturbative solution of (\ref{dimensionless wave}) we have
precisely the same interpretation: $\epsilon$ is the amplitude of
the leading order solution for $\Phi$ and our perturbative series is
only valid for small amplitude fluctuations.

Finally, we mention that to re-introduce dimensions correctly one
must multiplying (\ref{eq:perturbative series}) by $\M$:
\begin{equation}
    \phi = \phi_\KG + \sum_{i=1}^{\infty} \epsilon^{2i} \phi_{2i+1},
    , \quad \phi_\KG = \epsilon \M \Phi_1, \quad \quad \phi_{2i+1} = \epsilon \M
    \Phi_{2i + 1}.
\end{equation}
Note that $\phi_\KG$ satisfies the dimensionful Klein-Gordon
equation $(-\di_t^2 + \nabla^2)\phi_\KG = 0$.

\subsection{Planar symmetry}\label{sec:planar}

We now seek solutions of the perturbative equations of motion
(\ref{eq:perturbative EOMs}) with planar symmetry.  That is, if we
write the dimensionless spatial coordinates as $X^i = (X,Y,Z)$, then
all the fields $\Phi_i$ depend on $T$ and $X$ only.  Defining the
retarded and advanced times as
\begin{equation}
    U = T - X, \quad V = T + X,
\end{equation}
respectively, we assume that the leading order contribution $\Phi_1$
is a plane-fronted wave packet propagating in the positive $X$
direction:
\begin{equation}
    \Phi_1 = F(T-X) = F(U) ~.
\end{equation}
Here, $F$ is an arbitrary function.  As mentioned above, we can take
$\Phi_2 = 0$ without loss of generality.  For $\Phi_3$, we find that
(\ref{eq:third order}) reduces to
\begin{equation}
    \frac{\di^2 \Phi_3}{\di U \di V} = - \frac{1}{2} F''(U)
    [F'(U)]^2 ~,
\end{equation}
with general solution
\begin{equation}\label{GenSolPlaneWave}
    \Phi_3 = g_\text{L}(V) + g_\text{R}(U) -\frac{1}{6} V [F'(U)]^3 ~,
\end{equation}
where $g_\text{L}$ and $g_\text{R}$ are arbitrary functions.  To fix
these functions, we need to choose initial conditions.  A physically
interesting situation involves preparing the total field $\Phi$ such
that the polymer corrections are zero on some initial hypersurface
$\Sigma_0$; i.e., $\Phi$ corresponds to a wave packet that initially
evolves according to the Klein-Gordon equation. Obviously, the
particular choice of the initial data hypersurface will affect the
form of the solution.  For simplicity, we choose $\Sigma_0$ to be
the null hypersurface $V = 0$ and we require
\begin{equation}\label{eq:null initial condition}
    \Phi_3\big|_{V=0} = 0 ~.
\end{equation}
This means that $g_\text{L}(V) = 0$ and $g_\text{R}(U) = 0$.
Restoring dimensions, we obtain the following solution for $\phi$:
\begin{equation}\label{eq:phi series}
    \phi = M f \left[ 1 - \frac{v}{6f} \frac{M^2}{\M^4}
    \left(\frac{df}{du}\right)^3 + \cdots \right],
\end{equation}
where
\begin{equation}
    M = \epsilon \M, \quad u = t - x, \quad v = t+x, \quad f =  f(u) = F(\M u).
\end{equation}
Here, $M$ is a mass scale representing the amplitude of the leading
order (classical) solution.
We now consider a couple of explicit solutions which are of physical
interest. If the leading order term in the series (\ref{eq:phi
series}) represents a Gaussian wave packet $f = e^{-{\beta^2}
u^2}e^{iku}$ where the width of the wave-packet is $1/\beta$, we
obtain
\begin{equation}\label{eq:wave packet solution}
    \phi = M e^{-{\beta^2} u^2}e^{iku} \left[ 1 + \frac{1}{6} \frac{M^2 k^2}
    {\M^4} \left( 1 + \frac{2i{\beta^2} u}{k} \right)^3 ik v e^{-2{\beta^2} u^2}e^{2iku} + \cdots
    \right].
\end{equation}
Conversely, if the leading term represents a plane wave $f= e^{iku}
= e^{ik(t-x)}$ then we get
\begin{equation}\label{eq:plane wave solution}
    \phi = M e^{ik(t-x)} \left[ 1 + \frac{1}{6} \frac{M^2 k^2}
    {\M^4} ik(t+x) e^{2ik(t-x)} + \cdots \right].
\end{equation}
In either case, we see that the polymer corrections are small if
\begin{equation}
    kv = k (t + x) \ll \left(\frac{\M}{M}\right)^2 \left( \frac{\M}{k}
    \right)^2.
\end{equation}
Naively, we recover Klein-Gordon solutions in the limit of small
amplitudes $M \ll \M$ and/or long wavelengths $k \ll \M$. However,
notice that when $(t+x)$ becomes large, the leading order polymer
correction grows to become comparable to the Klein-Gordon
contribution. In other words, the perturbative solutions will
eventually break down. This implies that if we wait long enough,
then the nonlinear quantum effects will affect the wave propagation
significantly, irrespective of the values of $k$ or $M$.

Previously, we considered the initial data hypersurface to be null.
We now consider the situation where initial data are specified at
the $T = 0$ hypersurface. As earlier, we require ``purely
Klein-Gordon'' data on the initial hypersurface.  That is, $\Phi_3$
should satisfy the following conditions
\begin{equation}
    \Phi_3\big|_{T=0} = 0, \quad \dot\Phi_3\big|_{T=0} = 0.
\end{equation}
Here it is slightly more involved to find explicit expressions for
$g_\text{L}$ and $g_\text{R}$. In the case where the leading order
term represents a plane wave, we get
\begin{equation} \label{T0ICSolutions}
    \phi = M e^{ik(t-x)}\left[1 +  \frac{i}{3} \frac{k^2 M^2}{\M^4}
    e^{2ik(t-x)} kt \left( 1 - \frac{e^{-3ikt}\sin 3kt}{3kt} \right) + \cdots
\right].
\end{equation}
It follows from the expression (\ref{T0ICSolutions}) of perturbative
solutions that polymer corrections are small only if
\begin{equation}
    kt \ll \left(\frac{\M}{M}\right)^2 \left( \frac{\M}{k}
    \right)^2.
\end{equation}
This is effectively the same criterion as found earlier for the
initial condition specified at null hypersurface (\ref{eq:null
initial condition}). In particular, we see that after a sufficiently
long time the polymer corrections will become non-negligible and the
perturbative solutions will break down.

\subsection{Spherical symmetry}\label{sec:spherical}

It is also interesting to solve the perturbation equations
(\ref{eq:perturbative EOMs}) under the assumption of spherical
symmetry; i.e., the fields are functions of $t$ and a radial
coordinate $r$. If we assume that the leading order contribution
represents an outgoing spherical wave packet, we obtain the
following solution:
\begin{equation}\label{eq:spherical wave packet}
    \phi = \frac{Me^{-{\beta^2} u^2}e^{iku}}{kr} \left[ 1 + \frac{2i}{3}\frac{k^2 M^2}
    {\M^4} \left( \frac{1}{kv_0} - \frac{1}{kv} \right) \left( 1 +
    \frac{2i{\beta^2} u}{k} \right)^3 e^{-2{\beta^2} u^2} e^{2iku}
    + \cdots \right] ~.
\end{equation}
Here, $u = t - r$ and $v = t + r$.  We have imposed the boundary
conditions such that $\phi$ reduces to the Klein-Gordon solution
$\phi = Me^{iku}/kr$ on an initial null hypersurface $v = v_0$.   It
is interesting to note that for late times (i.e, $v \rightarrow
\infty$) the solution reduces to
\begin{equation}
    \phi = \frac{Me^{-{\beta^2} u^2}e^{iku}}{kr} \left[ 1 + \frac{2i}{3}\frac{k M^2}
    {v_0 \M^4} \left( 1 +
    \frac{2i{\beta^2} u}{k} \right)^3 e^{-2{\beta^2} u^2} e^{2iku}
    + \cdots \right] ~,
\end{equation}
which is actually a solution of the ordinary Klein-Gordon equation
$(-\di_t^2 + \nabla^2)\phi = 0$. In contrast to case of planar
waves, the polymer effects on spherical waves gradually decreases as
the wave packet travels away from the initial surface. The reason
behind this is that the effective amplitude of the Klein-Gordon part
of the solution decays as $1/v$ when it is far away from $r = 0$.
Since the polymer corrections are explicitly \emph{amplitude
dependent}, it follows that we recover Klein-Gordon behaviour as $v
\rightarrow \infty$. It may be emphasized that this behaviour
however does not mean that the asymptotic form of $\phi$ exactly
matches the Klein-Gordon prediction. In particular, the polymer
effects generate a whole series of additional wave packets with
frequency $3k$, $5k$, etc in addition to the original wave packet
with frequency $k$, The relative amplitude of the ``polymer
harmonic'' with $k_n = (2n+1)k$ will be roughly $(kM^2/v_0\M^4)^n$,
where $v_0$ can be taken as the initial radius of the wave packet.
From this, it follows that our perturbative solutions will be valid
at infinity if
\begin{equation}
    \frac{k}{\M} \ll (\M v_0) \left( \frac{\M}{M} \right)^2.
\end{equation}
That is, low wavenumber and low amplitude waves are well described
by perturbative techniques.

\section{Phenomenological implications}\label{sec:phenomenology}

In this section, we explore some phenomenological implications of
the perturbative solutions of the semi-classical Klein-Gordon
equation.  We begin by analyzing the propagation of a planar
wave-packet and derive its group velocity and an effective
``dispersion relation'' (which must be interpreted with care). We
then examine the conditions under which our perturbative solutions
break-down, and hence derive a ``coherence length'' over which
planar waves resemble solutions of the ordinary Klein-Gordon
equation.  Finally, we consider the spherical wave-packet solutions
in the context of polymer modifications to the spectrum of gamma ray
bursts.

\subsection{Group velocity of polymer wave-packets and effective dispersion
relations}\label{sec:dispersion}

As mentioned earlier, there has been much discussion in the
literature concerning ``modified dispersion relations'' of the form
(\ref{DSRDispersion}), especially in the context of doubly special
relativity \cite{AmelinoDSR,MSDSR}.  Such formulae are consistent
with linear wave equations with higher derivatives; for example, an
equation of the form $[-\di_t^2 + \nabla^2 - l_\pl^2 \nabla^4] \phi
= 0$ leads to the dispersion relation $\omega^2 = k^2(1 + l_\pl^2
k^2)$.  However, for nonlinear wave equations such as the
polymer-corrected Klein-Gordon equation (\ref{waveeqn}), the
derivation and interpretation of dispersion relations is more
complicated.  In this section, we suggest a physically-motivated
derivation of an effective dispersion relation based on the
propagation of spatially localized wave-packets.

Let us begin by considering the polymer corrected planar wave
solution (\ref{eq:phi series}).  We assume that the leading order
Klein-Gordon term in the series represents a spatially localized
waveform with characteristic frequency $k$:
\begin{equation}
    f(u) = h(ku) e^{iku}.
\end{equation}
The real and dimensionless function $h$ defines the envelope of the
wave-packet. For simplicity, we restrict ourselves to symmetric
wave-packets with $h(\xi) = h(-\xi)$ and define $M$ such that
$h(0)=1$. Also, we assume that $h$ vanishes strongly at infinity
such that integrals of the form $\int \xi^n h^2(\xi) d\xi$ are
finite. Sufficient conditions for this hold are that $h$ has compact
supports or decays exponentially for large arguments.  Note that our
choice of $f(u)$ means that the Klein-Gordon contribution to the
perturbation series $\phi_\KG = M h(ku) e^{iku}$ is self-similar;
i.e., $\phi_\KG$ is invariant under a rescaling of the coordinate $u
\rightarrow \ell u$ if we also rescale the wavenumber $k \rightarrow
k/\ell$.

In this subsection, we are interested in the leading order polymer
effects, so we truncate the series (\ref{eq:phi series}) as follows:
\begin{equation}
    \phi \approx M h(ku) e^{iku}  - \frac{v}{6} \frac{M^3}{\M^4}
    \left\{\frac{d}{du} \left[h(ku) e^{iku}\right] \right\}^3.
\end{equation}
We define the position of the packet at a given time $t$ by the
average value of $x$:
\begin{equation}\label{AvgPositionDef}
    x_\text{avg} = \frac{\int_{-\infty}^{\infty} dx ~x
    \phi^{*}\phi} {\int_{-\infty}^{\infty} dx ~\phi^{*}\phi} = t -
    \frac{\int du ~u \phi^{*}\phi} {\int du \phi^{*}\phi},
\end{equation}
where we have made use of $u = t - x$.  In general, $x_\text{avg}$
will depend on time and the wavenumber $k$ of the packet.  It is
easy to confirm that in the Klein-Gordon limit $\M \rightarrow
\infty$ we get $x_\text{avg} = t$; i.e. the pulse propagates at the
speed of light.

To calculate $x_\text{avg}$ in general, it useful to define
\begin{equation}
    \mathcal{Q}(\xi) = h(\xi)e^{i\xi} \left\{ \frac{d}{d\xi}
    \left[ h(\xi) e^{-i\xi}\right] \right\}^3 + \text{c.~c.}
\end{equation}
From the fact that $h$ is even, it follows that $\mathcal{Q}$ is and
odd function of $\xi$. We also define the dimensionless moments
\begin{equation}
 \label{NuNDef}
 \mu_n \equiv \int_{-\infty}^{\infty} d\xi ~\xi^n h^2(\xi), \quad
 \nu_n \equiv \int_{-\infty}^{\infty} d\xi ~\xi^n \mathcal{Q}(\xi) ~,
\end{equation}
with $n = 0,1,2\ldots$  From the parity of $h$ and $\mathcal{Q}$,
these satisfy $\mu_{2n+1} = 0$ and $\nu_{2n} = 0$.  Using these
definitions and the fact that $v = 2t-u$, we find that
\begin{equation}
    \phi \phi^* = M^2 |h(ku)|^2 - \frac{k^3(2t-u)}{6} \left(\frac{M}{\M}
    \right)^4 \mathcal{Q}(ku) + \mathcal{O}\left( \frac{1}{\M^8} \right),
\end{equation}
which leads to
\begin{equation}
    \int u^n \phi^* \phi \,du = \frac{M^2}{k^{n+1}} \mu_n - \frac{t}{3k^{n-2}} \left(
    \frac{M}{\M} \right)^4 \nu_n + \frac{1}{6k^{n-1}} \left( \frac{M}{\M}
    \right)^4 \nu_{n+1} + \mathcal{O}\left( \frac{1}{\M^8} \right).
\end{equation}
This gives the effective position of the pulse as a function of
time:
\begin{equation}
    x_\text{avg} = t \left( 1 + \frac{M^2 k^2}{3\M^4} \frac{\nu_1}{\mu_0}
    \right) + \mathcal{O}\left( \frac{1}{\M^8} \right).
\end{equation}
From this expression, we see that the polymer wave-packet travels
with a group velocity different from unity
\begin{equation}\label{eq:group velocity}
    v_g = \frac{dx_\text{avg}}{dt} = 1 + \frac{M^2 k^2}{\M^4}
    \mathcal{F}_h + \mathcal{O}\left( \frac{1}{\M^8}
    \right), \quad \mathcal{F}_h \equiv \frac{\nu_1}{3\mu_0}.
\end{equation}
Here, $\mathcal{F}_h$ is a dimensionless form factor that depends on
the wave-packet profile $h$ only.  For example, if $h$ is a Gaussian
we obtain
\begin{equation}
    h(\xi) = e^{-\xi^2/\alpha^2} \quad \Rightarrow \quad
    \mathcal{F}_h = \frac{\sqrt{2}}{96} \left( \frac{\alpha^2 - 12}{\alpha^2}
    \right) e^{-\alpha^2/4}.
\end{equation}
We see from (\ref{eq:group velocity}) that the polymer correction to
the group velocity depends not only of the wavenumber $k$, but also
on the Klein-Gordon amplitude $M$ and the shape of the pulse
$\mathcal{F}_h$.  This formula for $v_g$ can be physically
interpreted by noting that the characteristic energy density
associated with the Klein-Gordon part of the wave-packet is
\begin{equation}
    \rho_\KG = \frac{1}{2} \left[ (\di_t \phi_\KG )^2
    + (\nabla \phi_\KG)^2 \right] \sim M^2 k^2.
\end{equation}
Then we see that
\begin{equation}
    v_g - 1 = \mathcal{O} \left(\frac{\rho_\KG}{\M^4} \mathcal{F}_h\right).
\end{equation}
That is, the polymer correction to the group velocity scales like
the density of the wave-packet normalized by the ``polymer density''
$\M^4$.  This is somewhat akin to the situation in loop quantum
cosmology, where polymer corrections to the Friedmann equation scale
like the density normalized by the gravitational polymer scale
$M^4_\pl$.  Previously in the literature, the astrophysical and
experimental consequences of modified group velocities have been
derived by considering corrections that depend only on $k$
\cite{gamma-exp}.  This assumption was motivated by \emph{ad hoc}
dispersion relations of the form (\ref{DSRDispersion}). Here, we see
that the corrections can depend on the amplitude and shape of the
wave-packet as well. This crucial difference is due to the fact that
we are dealing with a nonlinear semi-classical wave equation.  It is
worthwhile stressing that the above group velocity has been derived,
not postulated as in previous work.

Given the above expression for the group velocity, one may be
tempted to derive an effective dispersion relation by setting $v_g =
d\omega/dk$ and integrating with respect to $k$.  This can be done,
and we find
\begin{equation}\label{dispersion relation}
    \omega^2 = k^2 \left[ 1 + \frac{2}{3}\frac{M^2 k^2}{\M^4}
    \mathcal{F}_h + \mathcal{O}\left(\frac{1}{\M^8}\right) \right].
\end{equation}
On first glance, this seems to be of the same form as
(\ref{DSRDispersion}) with $\gamma = 2$.  But we again note that
corrections depend on the amplitude $M$ and shape $\mathcal{F}_h$ of
the original wave-packet.  We comment that amplitude dependent
dispersion relations are a common consequence of nonlinear wave
equations; for example, they arise in the theory of deep water
surface waves \cite{wavebook}.

We conclude by noting that significant caution is warranted if we
try to interpret (\ref{dispersion relation}) as the dispersion
relation for some linear wave equation with higher derivative terms.
For example, the fact that $d^2\omega/dk^2 \ne 0$ would na\"{\i}vely
imply that our wave-packet is spreading as time progresses.  But an
explicit calculation of the effective width of the pulse gives
\begin{equation}
    \Delta x^2 \equiv \frac{\int_{-\infty}^{\infty} dx ~x^2
    \phi^{*}\phi} {\int_{-\infty}^{\infty} dx ~\phi^{*}\phi} -
    x_\text{avg}^2 = \frac{\mu_2}{k^2\mu_0} + \frac{1}{6M^2 \mu_0} \left( \frac{M}{\M} \right)^4 \left(
    \nu_3 - \frac{\mu_2\nu_1}{\mu_0}  \right) +
    \mathcal{O}\left( \frac{1}{\M^8} \right).
\end{equation}
Note this is time independent to leading order in $\M$. So although
the dispersion relation insinuates that the packet is spreading, a
direct calculation shows that this is not the case to leading order.
Fundamentally, simple dispersion relations of the form
(\ref{DSRDispersion}) are not capable of capturing all of the
interesting dynamics associated with our nonlinear wave equation
(\ref{waveeqn}).

\subsection{Effective coherence length for planar
waves}\label{sec:coherence}

In this subsection, we discuss the break-down of the perturbative
solutions to the polymerized wave equation and the associated
observational effects. As we have seen in Sections \ref{sec:planar}
and \ref{sec:spherical}, the polymer quantum effects accumulate or
dissipate with time for planar and spherical waves, respectively.
So, to maximize our chances of seeing deviations from conventional
theory we should look at highly collimated beams of radiation. To
quantify how big polymer effects are in such beams, we rewrite our
polymer-corrected plane wave solution (\ref{eq:plane wave solution})
as
\begin{equation}\label{eq:approx wave}
    \phi = \phi_\KG \left[ 1 + \frac{i}{6} \left(\frac{t+x}{d_\poly}
    \right)
    e^{2ik(t-x)} + \cdots \right],
\end{equation}
where
\begin{equation}
    \phi_\KG = Me^{ik(t-x)}, \quad d_\poly = t_\poly = \frac{1}{k}
    \left(\frac{\M}{M}\right)^2 \left( \frac{\M}{k}
    \right)^2 ~.
\end{equation}
As we have mentioned before, the magnitude of the polymer
corrections inside the square brackets becomes larger as time goes
on, and our perturbative solutions are valid for
\begin{equation}
    t \ll t_\poly ~.
\end{equation}
That is, when this condition is satisfied the approximation $\phi
\approx \phi_\KG$ holds.  We note that we can obtain a very similar
expression to (\ref{eq:approx wave}) if $\phi_\KG$ is a wave packet
whose width is much larger than $1/k$.  In that case, the validity
of the classical solution demands that the distance $d$ traveled by
the wave packet satisfy
\begin{equation}
    d \ll d_\poly ~.
\end{equation}
It is useful to think of $t_\poly$ and $d_\poly$ as a ``coherence
time'' and ``coherence length'', respectively, because they
represent the time or distance over which the polymer corrections to
a classical wave packet remain small.

More practical expressions for $t_\poly$ and $d_\poly$ come from
noting that the luminosity (particles per unit area per unit time)
associated with the Klein-Gordon part of the beam is
\begin{equation}
    L_\KG = \frac{\rho_\KG}{k} = \frac{1}{2k} \left[ (\di_t \phi_\KG )^2
    + (\nabla \phi_\KG)^2 \right] \sim M^2 k ~.
\end{equation}
Restoring conventional CGS units, this yields
\begin{equation}
    d_\poly = l_\pl \left( \frac{\M}{M_\pl} \right)^4 \left( \frac{E_\KG}{E_\pl} \right)^{-2}
    \left( \frac{L_\KG}{L_\pl} \right)^{-1}, \quad L_\pl = \frac{1}{l_\pl^2 t_\pl} =
    7\times 10^{108}\,\text{cm}^{-2}\,\text{sec}^{-1} ~,
\end{equation}
where $E_\KG = \hbar k /c$ is the energy per particle in the beam
and $L_\pl$ is the Planck luminosity. From this, we see that the
coherence length is minimized for high-energy and large luminosity
beams.

One of the most luminous artificial beams ever constructed is in the
Large Hadron Collider (LHC) \cite{Bruning}.  In that device, the
proton luminosity is $L_\KG \sim
10^{34}\,\text{cm}^{-2}\,\text{sec}^{-1}$ at an energy of $E_\KG
\sim 7\,\text{TeV}$.  This corresponds to
\begin{equation}
    d_\poly \sim \left(\frac{\M}{M_\pl}\right)^4
    10^{45}\,\text{Gpc} ~.
\end{equation}
The coherence length is longer if we consider the most energetic
cosmic rays: The Pierre Auger Observatory observed 561 particles of
energy $\sim 10^{10}\,\text{GeV}$ with an integrated exposure of
$\sim 10^{5}\,\text{km}^2\,\text{yr}$ \cite{Abraham}, which
corresponds to
\begin{equation}
    d_\poly \sim \left(\frac{\M}{M_\pl}\right)^4
    10^{84}\,\text{Gpc} ~.
\end{equation}
Unless $\M \ll M_\pl$, both of these distances are many orders of
magnitude larger than the size of the observable universe (which is
$\sim 3 \,\text{Gpc}$). Hence, if the polymer scale is of the same
order as the Planck scale, the Klein-Gordon equation gives an
excellent description of the physics in both scenarios, and polymer
effects will be very hard to observe. However, we should point out
that there is no \emph{a priori} reason to take $\M \sim M_\pl$ in
this model. Indeed if we take $\M = 5\,\text{TeV}$, we find that
$t_\poly = 1\,\text{sec}$ in the LHC. That is, polymer effects would
completely dominate the beam after only one second.  Since we are
unaware of any exotic beam behaviour in the LHC, this can serve as
an effective lower bound: $\M  \gtrsim 5\,\text{TeV}$.

\subsection{Spectral modifications to gamma ray
bursts}\label{sec:GRB}

We conclude this section by considering the spherical wave packets
of Section \ref{sec:spherical}. We can use these to model the
radiation emitted from a gamma-ray burst (GRB) if we assume that the
energy emitted by the progenitor is isotropic and we can approximate
the electromagnetic wave dynamics using a scalar field. Both
assumptions should be sufficient for the order of magnitude
estimates we seek here.

We imagine the following situation:  The gamma-ray progenitor is
located a distance of $d$ away from the earth.  At $t=t_0$ a
spherical pulse of radiation of radius $r_0 = t_0$ is emitted from
this progenitor.  The initial pulse profile is assumed to be
Gaussian:
\begin{equation}
    \phi \approx \phi_\KG = \frac{Me^{-{\beta^2}(t-r)^2}e^{ik(t-r)}}{kr},
    \quad (t\approx t_0, r \approx r_0) ~.
\end{equation}
Now, the signal $S(\tau)$ seen by an earth based detector will just
be the polymer solution (\ref{eq:spherical wave packet}) evaluated
at $r = d$:
\begin{equation}
    S(\tau) \equiv \phi\big|_{r=d} = \frac{Me^{-{\beta^2}\tau^2}e^{ik\tau}}{kd}
    \left[ 1 + \frac{i}{3}\frac{k M^2} {\M^4 r_0} \left( 1 +
    \frac{2i{\beta^2}\tau}{k} \right)^3 e^{-2{\beta^2}\tau^2} e^{2ik\tau}
    + \cdots \right] ~,
\end{equation}
with $\tau = t - d$ and assuming $d \gg r_0$.  The Fourier transform
of this signal is
\begin{equation}\label{eq:spectral profile}
    S(\omega) = \frac{1}{\sqrt{2\pi}}\int d\tau \,e^{-i\omega\tau}
    S(\tau) = S_1 e^{-(\omega-k)^2/4{\beta^2}} + i S_3 \left(\frac{\omega}{3k}\right)^3
    \left[1-\frac{18{\beta^2}}{\omega^2}\right] e^{-(\omega - 3k)^2/12{\beta^2}} + \cdots,
\end{equation}
where
\begin{equation}
    S_1 = \frac{M}{\sqrt{2}\beta kd},
    \quad S_3 = \frac{\sqrt{6}}{18}\frac{M^3}{\beta d r_0 \M^4}.
\end{equation}
The first term in (\ref{eq:spectral profile}) represents the
spectral line centered about $\omega = k$ that our detector would
have seen without polymer quantization.  The second term is the
result of the leading order polymer correction:  An additional
spectral line centered about $\omega = 3k$.  The relative amplitude
of the two lines is
\begin{equation}
    \frac{S_3}{S_1} \sim \frac{E_\tot k\beta}{r_0 \M^4},
\end{equation}
where we have noted that the total energy in the pulse initially is
\begin{equation}
    E_\tot = \int{\rho_\KG \, d^3x} \sim M^2/ \beta,
\end{equation}
assuming a long duration $k \gg \beta$ and a large initial radius
$r_0\beta \gg 1$.  Now, the initial pulse must have a size bigger
than the Schwarzschild radius associated with $E_\tot$; i.e., $r_0
\gtrsim E_\tot/M_\pl^2$, so we obtain an upper bound on $S_3/S_1$:
\begin{equation}
    \frac{S_3}{S_1} \lesssim \left( \frac{E_\KG}{E_\pl} \right) \left( \frac{\Delta
    t}{t_\pl} \right)^{-1} \left( \frac{\M}{M_\pl} \right)^{-4} \sim
    10^{-66} \left( \frac{E_\KG}{10^5\,\text{eV}} \right) \left( \frac{\Delta
    t}{1\,\text{sec}} \right)^{-1} \left( \frac{\M}{M_\pl} \right)^{-4},
\end{equation}
where $\Delta t$ is the duration of the pulse observed on earth and
$E_\KG$ is the energy per particle (photon) in the beam; for a GRB,
one typically has $\Delta t \sim 1\,\text{sec}$ and $E_\KG \sim
10^5\,\text{eV}$ \cite{Piran:2004}.  As in the case for the LHC, we
know of no evidence for exotic polymer effect in observed GRBs,
which suggests that $S_3/S_1 \ll 1$.  This yields another bound: $\M
\gtrsim 1\,\text{TeV}$.  This is virtually the same numeric result
we obtained from the LHC.

\section{Discussion}\label{sec:discussion}

We have developed an approach for studying polymer quantum field
theory on a curved background spacetime. The quantization procedure
is motivated by ideas used in loop quantum gravity. However, the
representation used here is significantly different to the one used
earlier \cite{als-fock}.  The polymer quantization differs from the
usual approach in two important aspects: the inner product is
background independent, and there is a built-in  mass (or length)
scale $\M$ due to the choice of basic observables.  Using
semi-classical methods we have derived an effective nonlinear wave
equation governing the field dynamics.  The form of this
semi-classical wave equation necessarily depends on the underlying
quantum state, but we explicitly find that it reduces to the
conventional Klein-Gordon equation at low energies.

In \S\ref{sec:perturbative}, we developed perturbative solutions to
the nonlinear wave equation.  The natural expansion parameter
involves the amplitude of the lowest order contribution. We then
specialized to waves with planar and spherical symmetry in
\S\ref{sec:planar} and \S\ref{sec:spherical}, respectively.  For the
plane-symmetric case, we found that the polymer corrections to the
classical solution actually grow in time. This suggests that if one
waits long enough, nonlinear effects will completely dominate the
Klein-Gordon contribution to the wave.

The converse is true for spherical waves, which tend to decrease in
amplitude as they travel outward from the center of symmetry. This
implies that the polymer effects remain bounded as time progresses.
This qualitative difference between planar and spherical wave
highlights one of the most important features of the model:
corrections to classical dynamics depend on more than just the
frequency of the wave.  Furthermore, if one begins with a spatially
localized spherical wave-packet with characteristic wavenumber $k$,
one obtains a whole series of wave-packets with wavenumber $k$,
$3k$, $5k$ \ldots at infinity.

In \S\ref{sec:phenomenology}, we derived some phenomenological
implications of the effective wave equation.  We found that the
group velocity of planar wave-packets depends on their frequency,
amplitude and shape.  This is in sharp contrast to many models in
the literature, which assume that the group velocity only recieves
frequency-dependent modifications.  From the these group velocities,
we derive effective dispersion relations for the model.  These bear
some resemblance to the dispersion relations of weakly nonlinear
waters waves in that they depend explicitly on amplitude. Our method
of extracting a dispersion relation  appears to be new, and has
potential applications to other  nonlinear wave equations.

The fact that our perturbative solution breaks down after a finite
amount of time has interesting practical implications, as discussed
in \S\ref{sec:coherence}.  This implies that there is a ``polymer
coherence length'' $d_\poly$ associated with plane-symmetric wave
packets.  After a wave packet has traveled a distance $d_\poly$,
polymer effects will completely dominate the classical behaviour. We
express $d_\poly$ as a function of the frequency and luminosity of
the beam.  We calculate $d_\poly$ for the proton-antiproton beam in
the Large Hadron Collider, and use the fact that no exotic behaviour
has been reported in that device to deduce a lower limit of $\M
\gtrsim 1 \, \text{TeV}$.  We also calculated $d_\poly$ for a highly
collimated beam of cosmic rays, and found that it is many of orders
of magnitude larger than the size of the observable universe.
Finally, we considered spherical gamma ray bursts in \S\ref{sec:GRB}
and demonstrated that if the characteristic frequency of the burst
is $\omega = k$, an earth-based observer will see additional
spectral lines centered at $\omega = (2n+1)k$.  From the amplitude
ratio of the classical to polymer spectral features, we deduced that
$\M \gtrsim 1\,\text{TeV}$ consistent with the LHC result mentioned
above.

It is interesting to note that all of the modifications to wave
propagation mentioned above are derived without including any
quantum gravity effects, unlike other approaches such as doubly
special relativity, loop quantum gravity, or non-commutative
spacetime models.  Another important feature worth emphasizing is
that fundamental discreteness in this model leads to nonlinearities
at short distances.  This is contrary to the common assumption that
the effects of fundamental discreteness manifest themselves via a
linear wave equation with higher derivative terms.

There are a number of possible developments based on the ideas we
have discussed, such as applications to Hawking radiation and to the
spectrum of fluctuations in inflationary cosmology, which have as
their basis a wave equation on a curved background. These are
presently being studied.  Another application is to homogeneous
scalar field propagation on an FRW background. This work provides
some novel results, including cosmological singularity avoidance and
a mechanism for inflation \cite{polyfrw}. The former result is a
concrete realization of the intuition that in the Hamiltonian
constraint, the compactification of the matter energy density that
results from the semi-classical effective Hamiltonian leads to a
bound on the gravitational terms in this constraint; the converse
occurs in the LQC  case, where it is the gravitational kinetic terms
that are compactified by polymer geometry.  We also show in
\cite{polyfrw} that the energy scale of polymer inflation is fixed
by $\M$, which leads to an entirely different way of constraining
the fundamental discreteness scale in our model.

\begin{acknowledgements}

This work was supported in part by the Natural Science and
Engineering Research  Council of Canada, and the Atlantic
Association for Research  in the Mathematical Sciences.

\end{acknowledgements}

\appendix

\section{The semi-classical wave equation on a curved background}
\label{app:wave equation on a curved background}

The method followed to derive an effective wave equation on
Minkowski spacetime is applicable also to an arbitrary background.
Here we restrict attention to metrics  with lapse $N=1$ and shift
$N^a=0$, and denote the spatial metric by $q_{ab}$ and its
determinant by $q$.

 Let $\mathcal{H}(P_\phi(x),\phi(x))$ be the Hamiltonian density and
define the effective Hamiltonian density $\mathcal{H}^{\rm eff}$ by
the expectation value
\be \mathcal{H}^{\rm eff} (P_0(x_k),\phi_0(x_k);\sigma,\lambda)
\equiv \frac{1}{2 }\ \langle P_0,\phi_0| {( \hat{P}_\phi^\lambda) ^2
\over {\sqrt{q}}}+ \sqrt{q}\ \widehat{\nabla^2\phi}   |P_0,\phi_0
\rangle_{x_k}^{\sigma,\lambda} ~. \ee
The semi-classical states given above have to be modified to take
into into the density weights, and the fact that that the peaking
value for the scalar field is actually for $\phi_f$ rather than
$\phi$.  With these changes,  to leading order in the width
$\sigma$, the kinetic part of the operator gives
\bea \langle P_0,\phi_0|  (\hat{P}_\phi^\lambda)^2  |P_0,\phi_0
\rangle_{x_k}^{\sigma,\lambda} &=& {q\over  \lambda^2}\ \langle
P_0,\phi_0|  \left( 2- \hat{U}_{2\lambda} (P_\phi(x_k))
 -  \hat{U}_{2\lambda}^\dagger (P_\phi(x_k)) \right)
|P_0,\phi_0 \rangle_{x_k}^{\sigma,\lambda}\nn\\
&=&  {q\over \lambda^2}\ \sin^2\left(  {P_0 \lambda\over \sqrt{q}  }
\right) + {\cal O}\left(\frac{1}{\sigma}\right), \eea
where the last term represents the width corrections of the
semi-classical states.
The scalar gradient operator  is diagonal in the basis, and as for
the flat space case, its discrete aspect can be ignored for
wavelengths large compared to the spatial lattice spacing
$\epsilon$. Thus  we can write
\be \mathcal{H}^{\rm eff} (x) = \frac{\sqrt{q}}{2\lambda^2}
\sin^2\left({P\lambda\over \sqrt{q}}\right) + \frac{\sqrt{q}}{2}\
\nabla^2\phi. \ee
 This Hamiltonian density leads to the modified wave equation
\be
 \frac{\di^2\phi}{\di t^2} - \left[ 1-4\lambda^2\left( \frac{\di\phi}{\di t} \right)^2
    \right]^{1/2}  \left[ \frac{1}{\sqrt{q} }\nabla^2 \phi - {1\over 4\lambda q}
\left(\frac{\di q}{\di t} \right) \arcsin \left(2\lambda
\frac{\di\phi}{\di t} \right) \right] = 0. \ee
%

%%%%%%%%%%   References  %%%%%%%%%%%


\begin{thebibliography}{99}
%


\bibitem{lqg-wave} R. Gambini, J. Pullin,  Phys. Rev. D59, 124021 (1999);    J. Alfaro  , Hugo A. Morales-Tecotl, Luis F. Urrutia, Phys. Rev. D65 (2002) 103509 (arXiv:hep-th/0108061);
H. Sahlmann, T. Thiemann, Class. Quant. Grav. 23 (2006) 909-954
(arXiv:gr-qc/0207031); C. N. Kozameh, M.F. Parisi  Class. Quant.
Grav. 21 (2004) 2617-2621 (arXiv:gr-qc/0310014).

\bibitem{myers-posp} Robert C. Myers, M. Pospelov, Phys. Rev. Lett. 90, 211601 (2003) 
 (hep-ph/0301124).

\bibitem{collins} J. Collins, A. Perez, D.Sudarsky, L. Urrutia, H. Vucetich  Phys. Rev. Lett. 93: 191301 (2004); e-Print: gr-qc/0403053.

\bibitem{unruh-disp} W. G. Unruh, Phys. Rev. D51, 2827-2838 (1995).

\bibitem{ted-disp} S. Corley, T. Jacobson, Phys. Rev. D54, 1568 (1996).

\bibitem{brand-disp} R. H. Brandenberger, J. Martin,  Phys. Rev. D63, 123501 (2001).

\bibitem{gamma-exp}  G. Amelino-Camelia, John Ellis, N.E. Mavromatos, D.V. Nanopoulos, Subir Sarkar, {\it Potential Sensitivity of Gamma-Ray Burster Observations to Wave Dispersion in Vacuo},
 Nature 393, 763 (1998).

\bibitem{GZK} Kenneth Greisen, Phys. Rev. Lett.16, 748 (1966); G.T. Zatsepin and V.A. Kuz�min, JETP Lett. 41 (1966) 78.

 \bibitem{ted-review}  T. Jacobson, S. Liberati, D. Mattingly, {\it Quantum gravity phenomenology and Lorentz violation},  Springer Proc. Phys. 98, 83 (2005)  (gr-qc/0404067);
 Annals Phys. 32, 150(2006) (astro-ph/0505267).

\bibitem{Thiemann:1997rt}
  T.~Thiemann,
  ``QSD V: Quantum gravity as the natural regulator of matter quantum field
  theories,''
  Class.\ Quant.\ Grav.\  {\bf 15}, 1281 (1998)
  [arXiv:gr-qc/9705019].


  \bibitem{qsd-n} T. Thiemann,  Class. Quant. Grav.15,  839 (1998).

 \bibitem{als-fock} A. Ashtekar. J. Lewandowski, H. Sahlmann, Class. Quant. Grav. 20 (2003) L11-1,
 (arXiv:gr-qc/0211012).


 \bibitem{hw-sing} V. Husain, O. Winkler, Class. Quant. Grav. 22, L 127 (2005); 
 Class. Quant. Grav. 22, L135 (2005). 


\bibitem{Thiemann:2001yy}
  T.~Thiemann,
  ``Introduction to modern canonical quantum general relativity,''
  arXiv:gr-qc/0110034.


\bibitem{Ashtekar:2004eh}
  A.~Ashtekar, J.~Lewandowski,
  ``Background independent quantum gravity: A status report,''
  Class.\ Quant.\ Grav.\  {\bf 21}, R53 (2004)
  [arXiv:gr-qc/0404018].

\bibitem{Smolin:2004sx}
  L.~Smolin,
  ``An invitation to loop quantum gravity,''
  arXiv:hep-th/0408048.

\bibitem{Rovelli:2008zza}
  C.~Rovelli,
  ``Loop quantum gravity,''
  Living Rev.\ Rel.\  {\bf 11} (2008) 5.


\bibitem{polymer}
%\bibitem{Ashtekar:2002sn}
  A.~Ashtekar, S.~Fairhurst and J.~L.~Willis,
  ``Quantum gravity, shadow states, and quantum mechanics,''
  Class.\ Quant.\ Grav.\  {\bf 20}, 1031 (2003)
  [arXiv:gr-qc/0207106].

\bibitem{halvorson} H. Halvorson,  ``Complementarity of representations in quantum mechanics'',
Stud. Hist. Phil. Mod. Phys., 35 (2004); arXiv: quant-ph/0110102.


\bibitem{hw-cosm} V. Husain, O. Winkler, Phys. Rev. D75, 024014 (2007).
e-Print: gr-qc/0607097.

 \bibitem{improved-dyn}
  A.~Ashtekar, T.~Pawlowski and P.~Singh,
  ``Quantum nature of the big bang: Improved dynamics,''
  Phys.\ Rev.\  D {\bf 74}, 084003 (2006)
  [arXiv:gr-qc/0607039].

\bibitem{Bojowald:2008zzb}
  M.~Bojowald,
  ``Loop quantum cosmology,''
  Living Rev.\ Rel.\  {\bf 11}, 4 (2008).

 \bibitem{polyfrw}
   G.~M.~Hossain, V.~Husain and S.~S.~Seahra,
  %``Non-singular inflationary universe from polymer matter,''
  arXiv:0906.2798 [astro-ph.CO].
  %%CITATION = ARXIV:0906.2798;%%

\bibitem{Bojowald:2001xe}
  M.~Bojowald,
  ``Absence of singularity in loop quantum cosmology,''
  Phys.\ Rev.\ Lett.\  {\bf 86}, 5227 (2001)
  [arXiv:gr-qc/0102069].

\bibitem{Date:2004fj}
  G.~Date and G.~M.~Hossain,
  ``Genericness of big bounce in isotropic loop quantum cosmology,''
  Phys.\ Rev.\ Lett.\  {\bf 94}, 011302 (2005)
  [arXiv:gr-qc/0407074].

\bibitem{AmelinoDSR}
  G.~Amelino-Camelia,
  %``Relativity in space-times with short-distance structure governed by an
  %observer-independent (Planckian) length scale,''
  Int.\ J.\ Mod.\ Phys.\  D {\bf 11}, 35 (2002)
  [arXiv:gr-qc/0012051].
  %%CITATION = IMPAE,D11,35;%%

\bibitem{MSDSR}
  J.~Magueijo and L.~Smolin,
  %``Lorentz invariance with an invariant energy scale,''
  Phys.\ Rev.\ Lett.\  {\bf 88}, 190403 (2002)
  [arXiv:hep-th/0112090].
  %%CITATION = PRLTA,88,190403;%%

\bibitem{wavebook}
    A.~M.~Kamchatnov, \emph{Nonlinear periodic wave and their
    modulations: An introductory course} (World Scientific, 2000)

\bibitem{Bruning}
  O.~S.~Bruning and P.~Collier,
  %``Building A Behemoth,''
  Nature {\bf 448}, 285 (2007).

\bibitem{Abraham}
  J.~Abraham {\it et al.}  [Pierre Auger Collaboration],
  %``Observation of the suppression of the flux of cosmic rays above $4\times
  %10^{19}$eV,''
  Phys.\ Rev.\ Lett.\  {\bf 101}, 061101 (2008)
  [arXiv:0806.4302 [astro-ph]].

\bibitem{Piran:2004}
  T.~Piran,
  %``The physics of gamma-ray bursts,''
  Rev.\ Mod.\ Phys.\  {\bf 76}, 1143 (2004)
  [arXiv:astro-ph/0405503].

\end{thebibliography}
\end{document}